\documentclass[12pt]{article}

\usepackage{standalone}

\usepackage{amsmath}
\usepackage{amsfonts}
\usepackage{amssymb}
\usepackage{amsthm}
\usepackage[T1]{fontenc}

\usepackage{algorithm}
\usepackage{algpseudocode}

\theoremstyle{definition} 

\usepackage[numbers]{natbib}
\usepackage[disable]{todonotes}
\usepackage{hyperref}

\usepackage{graphicx}
\graphicspath{{fig/}{}}

\usepackage{csvsimple}

\usepackage{arydshln} 

\title{Parameter identification for predator-prey system with sparse data}
\author{Eduard Campillo-Funollet, James Van Yperen} 

\date{\today}

\newcommand{\ta}{\tilde\alpha}

\newcommand{\wg}{\omega_\gamma}
\newcommand{\wb}{\omega_\beta}
\newcommand{\hwg}{\hat{\omega}_\gamma}
\newcommand{\hwb}{\hat{\omega}_\beta}
\newcommand{\vX}{\vec{X}}
\newcommand{\vY}{\vec{Y}}

\DeclareMathOperator{\diag}{diag}

\begin{document}

\maketitle

\section{Introduction}

Parameter identification from observations of dynamical parameters is a
fundamental problem in population biology. Dynamical systems provide powerful
mechanistic models of ecological systems, but optimisation methods often
require accurate initial guesses or fine tuning of hyperparameters to guarantee
convergence. In ecological applications, datasets are often subject to
considerable observation noise, and data is collected at sparse time points,
for example yearly census data for a population \cite{todman23}.

There are two computational bottlenecks for the parameter identification
problem. First, with sparse data the likelihood becomes more irregular, and
optimisation methods such as gradient ascent and BFGS struggle to converge, for
example due to extreme gradients pushing the method out of the convergence basin,
or due to requiring very short steps, and in consequence a large number of
iterations to avoid numerical issues. 

The second bottleneck is the stability of the ODE solver. Even if the
optimisation algorithm is converging, it may cross regions of the parameter
space that make the ODE stiff or numerical unstable in general. In the best
case this just leads to long running times, but often it creates runtime errors
to stop the method altogether, e.g. due to numerical overflows.

We present a computational framework for parameter identification in the
classic Lotka-Volterra predator-prey system that addresses these numerical
instabilities. To overcome the limitations of standard gradient methods, we
employ Natural Gradient Ascent (NGA) \cite{martens20}, leveraging the Fisher Information to
obtain a more stable optimization trajectory, in less iterations. We exploit
the non-dimensionalisation of the ODE to reduce the number parameters involved
in the dynamical system, treating the scaling factors as nuisance parameters to
reduce the dimensionality of the optimisation problem. Furthermore, we exploit
the properties of the predator-prey system to implement an efficient solver to
avoid the ODE solver step becoming too small in regions with very large
derivatives.

Let $x$ and $y$ represent the population of prey and predator respectively, subject to the classical Lotka--Volterra predator-prey dynamics \cite{lotka26,volterra26} given by the following system of ordinary differential equations
\begin{align}
x' &= \alpha x - \beta xy, \label{eqn:x}\\
y' &= \gamma xy - \delta y, \label{eqn:y}
\end{align}
with initial conditions $x(t_0) = x_0$ and $y(t_0) = y_0$ respectively. Here $\alpha$ represents the average growth rate of the prey, $\beta$ represents the average death rate of the prey due to predation from the predator, $\gamma$ represents the average growth rate of the predator due to predation of the prey, and $\delta$ represents the average death rate of the predator. Since we are considering predator-prey dynamics (and not mutualism, for example), all parameters and initial conditions are strictly positive except $\alpha$. A negative $\alpha$ implies that the prey population average death rate not associated to the predator is larger than the average birth rate. 

\section{Statistical Framework}

\subsection{Data Generation Process}

We assume that we have a discrete series of observations of some prey and population counts, denoted by $X_i$ and $Y_i$ respectively, at times $t_i$, for $i=0,\dots,n$. We let $\vX$ and $\vY$ denote the collection of each dataset. We assume that each datapoint $X_i$ and $Y_i$ are random observations from a probability distribution with mean $x(t_i)$ and $y(t_i)$ respectively. In this paper we consider
\begin{equation}
    \label{data_generating_process}
    \begin{aligned}
        X_i &\sim \mathrm{Poisson}(x(t_i)), \\
        Y_i &\sim \mathrm{Poisson}(y(t_i)),
    \end{aligned}
\end{equation}
for each $i=0,\dots,n$, but other probability distributions are used commonly in practice such as using a normal distribution or a negative binomial distribution. More generally, this corresponds to a setting with independent but not identically distributed samples. We let $f_{X_i}$ and $f_{Y_i}$ denote the probability mass functions associated to the observation random variables $X_i$ and $Y_i$ respectively. 

Given initial conditions $x_0$ and $y_0$, and parameters $\alpha$, $\beta$, $\gamma$ and $\delta$, the dynamics of $x(t)$ and $y(t)$ can be fully quantified via the solution to the ODE system \eqref{eqn:x}--\eqref{eqn:y}. This means that the observation random variables $X_i$ and $Y_i$ are actually parametric distributions defined by the parameters $\alpha$, $\beta$, $\gamma$ and $\delta$ and initial conditions $x_0$ and $y_0$. To be clear, we demonstrate this in the following way
\begin{equation}
    \label{data_generating_process_p}
    \begin{aligned}
    X_i &\sim \mathrm{Poisson}(x(t_i;\alpha,\beta,\gamma,\delta,x_0,y_0)), \\
    Y_i &\sim \mathrm{Poisson}(y(t_i;\alpha,\beta,\gamma,\delta,x_0,y_0)).
    \end{aligned}
\end{equation}
Thus, we look to obtain parameters $\alpha$, $\beta$, $\gamma$ and $\delta$ and initial conditions $x_0$ and $y_0$ that maximises the likelihood defined by the observation random variables $\vX$ and $\vY$, namely 
\begin{equation}
    \label{likelihood}
    \ell\left(\alpha,\beta,\gamma,\delta,x_0,y_0;\vX,\vY\right) = \ell_{\vX}\left(\alpha,\beta,\gamma,\delta,x_0,y_0\right) \ell_{\vY}\left(\alpha,\beta,\gamma,\delta,x_0,y_0\right)
\end{equation}
where
\begin{equation*}
    \ell_{\vX}\left(\alpha,\beta,\gamma,\delta,x_0,y_0\right) = \prod_{i=0}^n f_{X_i}\left(x\left(t_i;\alpha,\beta,\gamma,\delta,x_0,y_0\right)\right)
\end{equation*}
and
\begin{equation*}
     \ell_{\vY}\left(\alpha,\beta,\gamma,\delta,x_0,y_0\right) = \prod_{i=0}^n f_{Y_i}\left(y\left(t_i;\alpha,\beta,\gamma,\delta,x_0,y_0\right)\right).
\end{equation*}

\subsection{Dynamical vs statistical parameters}

A non-dimensionalisation of the system \eqref{eqn:x}--\eqref{eqn:y} naturally considers some of these parameters as scaling factors. This step is common in the analysis of differential equations, and it is important for numerical stability. Here, we will exploit it to facilitate parameter estimation by separating the estimation of parameters that play a role in the dynamical system, from parameters that can be seen as scale parameter from the statistical observation model. Let $s = \delta t$, then we use the following non-dimensionalisation 
\begin{align}
    u(s) &= \log\left(\frac{\gamma}{\delta} x(t)\right),\label{eqn:u_def}\\
    v(s) &= \log\left(\frac{\beta}{\delta} y(t)\right).\label{eqn:v_def}
\end{align}
The scaled variables $u$ and $v$ satisfy the following ordinary differential equation
\begin{align}
    \dot{u} &= \ta - e^v,\label{eqn:u}\\
    \dot{v} &= e^u - 1,\label{eqn:v} 
\end{align}
with initial conditions $u(s_0) = u_0$ and $v(s_0) = v_0$ respectively, where we set $\ta=\alpha/\delta$ and the dot above the variables denotes the derivative with respect to $s$. This demonstrates that the dynamics of the ODE can be controlled by $\ta$ and the initial conditions $u_0$ and $v_0$, whilst $\delta$, $\beta$ and $\gamma$ control the mapping from the ODE to the data. Since the mapping from $(\alpha,\beta,\gamma,\delta,x_0,y_0)$ to $(\ta, \wb,\wg,\delta,u_0,v_0)$ is bijective, where we have set $\wb = \delta/\beta$ and $\wg =\delta/\gamma$, we can rewrite the likelihood \eqref{likelihood} in terms of the new parameters, namely we have
\begin{equation*}
    \ell_{\vX}\left(\alpha,\beta,\gamma,\delta,x_0,y_0\right) \equiv \ell_{\vX}\left(\ta,\wg,\delta,u_0,v_0\right) = \prod_{i=0}^n f_{X_i}\left(\wg \exp\left(u\left(\delta t_i;\ta,u_0,v_0\right)\right)\right)
\end{equation*}
and
\begin{equation*}
    \ell_{\vX}\left(\alpha,\beta,\gamma,\delta,x_0,y_0\right) \equiv \ell_{\vY}\left(\ta,\wb\delta,u_0,v_0\right) = \prod_{i=0}^n f_{Y_i}\left(\wb \exp\left(v\left(\delta t_i;\ta,u_0,v_0\right)\right)\right).
\end{equation*}
It is important to note that the likelihood for $\vX$ does not depend on $\wb$ and the likelihood for $\vY$ does not depend on $\wg$. 

\subsection{MLE for the scaling parameters}
Calculating Maximum Likelihood Estimators for parameters in a standard statistical approach that enables a data-driven formula to be used to estimate parameters, rather than having to use a numerical optimiser. Typically, for parameter estimation of ODEs, none of the parameters have explicit MLE expressions due to the dependence of the ODE solution on the parameter. However, in this case, we have shown above two important things: the first is that $\wg$ and $\wb$ only scale the ODE solution and thus the ODE solution does not depend on them, and secondly the individual likelihood functions only depend on one each of $\wg$ and $\wb$. These enable us to obtain explicit MLE expressions to $\wg$ and $\wb$ in terms of the ODE solution and the other parameters and initial conditions. Namely, since we assuming our data is generated using a Poisson distribution, for $\wg$ we have
\begin{align*}
    \partial_{\wg} \log \ell_{\vX}\left(\ta,\wg,\delta,u_0,v_0\right) = \sum_{i=0}^n \left(\frac{X_i}{\wg} - \exp\left(u\left(\delta t_i;\ta,u_0,v_0\right)\right)\right),
\end{align*}
which, when set to 0, gives us
\begin{align*}
    \hwg = \frac{\sum\limits_{i=0}^n X_i}{\sum_{i=0}^n \exp\left(u\left(\delta t_i;\ta,u_0,v_0\right)\right)}.
\end{align*}
Doing the same for $\wb$ gives
\begin{align*}
    \hwb = \frac{\sum\limits_{i=0}^n Y_i}{\sum_{i=0}^n \exp\left(v\left(\delta t_i;\ta,u_0,v_0\right)\right)}.
\end{align*}
With MLE expressions at hand for $\wb$ and $\wg$, we can treat them as nuance parameters: we focus on estimating $\ta,\delta,u_0$ and $v_0$, and we use the above formulas to compute the corresponding MLE for $\hwb$ and $\hwg$. This means that we only need to use a numerical optimisation algorithm to estimate four parameters rather than six, which has two advantages: typically the optimiser requires less function evaluations, and the convergence basin for the optimiser is larger, requiring less accurate initial guesses. See the work by \cite{campillo26} for an example of a compartmental model where estimating four parameters is more stable than estimating six.

\section{Efficient numerical solution of the ODE}
\begin{figure}[!htb]
\includegraphics[width=0.94\linewidth]{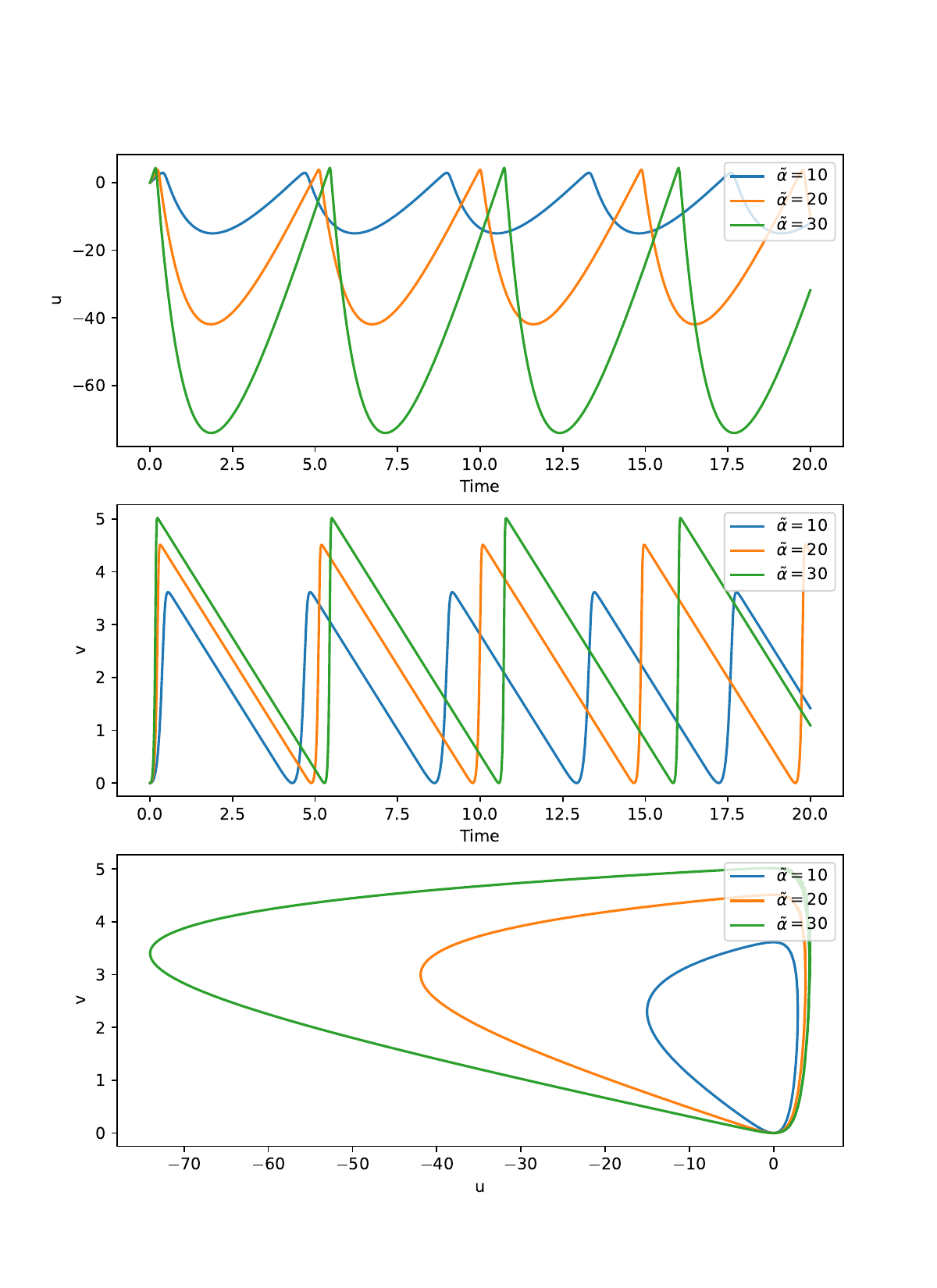}
\caption{Solution to \eqref{eqn:u}--\eqref{eqn:v} for a range of parameters $\ta$. Initial conditions are set to $(u_0,v_0)=(0,0)$.}\label{fig:uv_examples}
\end{figure}

For parameter estimation of ODEs, a numerical optimisation algorithm will have to evaluate the solution of the ODE many times, which means we need to numerically solve the ODE many times. The numerical solution to ODEs is a large field that contains many prebuilt solvers that users of many programming languages can use. Typically, these solvers are either adaptive, where the time step is changed based on local properties of the solution and forcing, or non-adaptive, where the time step is fixed but the numerical scheme is usually extensive and has high-order convergence properties. Typically, when these prebuilt solvers and employed, the choice of parameters lead to reasonable dynamics in the solution. However, when doing parameter estimation, the numerical optimiser might end up searching in the parameter space where the dynamics of the ODE solution become erratic. This is troublesome because it means that non-adaptive methods can accrue large errors quickly, and adaptive solvers can take an exceedingly long time as the adaptive time step is sent towards 0 to capture the erratic behaviour. This is precisely what happens in our case for large values of $\ta$, as can be seen in Figure \ref{fig:uv_examples}. Whilst $u$ behaves reasonably nicely, $v$ changes population size very dramatically. Mathematically this implies that $\dot{v}$ is very large over a small timescale. In the phase plane, we can see that large $\ta$ causes the predator-prey cycles flatten against the axes. To bypasses the issue of a tiny time step when utilising an adaptive solver, we have produced a novel method that uses properties of this specific ODE system.  

The key observation is that the derivatives of $u$ and $v$ become large \textit{at different times}. This is due to the lag between the prey and predator cycles: the peak of the predator follows the peak of the prey after some non-zero time. We can rewrite the system of equations as two independent second order ODEs, each equation depending only on $u$ or $v$. Then, the adaptive time step is based purely on one of the solutions, and when it shrinks below a certain threshold we can switch to solving the ODE associated to the other solution. The details can be found in Algorithm \ref{ode_solver}. We note that now we are numerically solving four first-order ODEs rather than two, this is however not a computational problem and is vastly outweighed by the benefits of the adaptive approach. We have also implemented this scheme na{\"i}vely using prebuilt solvers, there could be further numerical benefits if it is implemented differently however this is not the scope of this manuscript. 

To derive the second order equations for $u$ and $v$, we take a derivative in \eqref{eqn:u} (resp. \eqref{eqn:v}) and substitute $\dot{v}$ (resp. $\dot{u}$) using \eqref{eqn:v} (resp. \eqref{eqn:u}) to obtain
\begin{align}
\ddot{u} = -(e^u -1)(\ta - \dot{u}),\label{eqn:u2}\\
\ddot{v} = (\ta - e^v)(\dot{v}+1).\label{eqn:v2}
\end{align}
We note that we can obtain the initial conditions for $\dot{u}$ and $\dot{v}$ by using the ODE, namely $\dot{u}_0 = \ta - e^{v_0}$ and $\dot{v}_0 = e^{u_0} - 1$. We can then solve one of these second order ODEs and, by rearranging \eqref{eqn:u} or \eqref{eqn:v}, use the solution to obtain the other variable. More precisely,
\begin{align*}
u &= \log(\dot{v}+1),\\
v &= \log(\ta - \dot{u}).
\end{align*}

\begin{algorithm}[!htb]
\caption{Algorithm depicting the solver that utilises switching numerically solving for $u$ and for $v$.}
\label{ode_solver}
\begin{algorithmic}[1]
\Require Current states $u_0$, $v_0$, and parameter $\ta$
\State Compute $\dot{u}_0 = \ta-e^{v_0}$; $\dot{v}_0=e^{u_0} - 1$.
\State Compute the next adaptive steps for each solver: $h_u(u_0,\dot{u}_0)$, $h_v(v_0,\dot{v}_0)$.
\If{$h_u > h_v$}
    \State Compute $u_1$ and $\dot{u}_1$ by numerically solving \eqref{eqn:u2} with step $h_u$.
    \State Compute $v_1 = \log(\ta - \dot{u}_1)$.
\Else 
    \State Compute $v_1$ and $\dot{v}_1$ by numerically solving \eqref{eqn:v2} with step $h_v$.
    \State Compute $u_1 = \log(\dot{v}_1 + 1)$.
\EndIf
\end{algorithmic}
\end{algorithm}

\section{Gradient ascent methods}

Since the estimation problem for $\delta$, $\ta$, $u_0$ and $v_0$ is non-linear, we will rely on a numerical optimisation scheme to obtain the MLE for these parameters. There are two key performance measures: stability and convergence. Here stability is concerned with whether the numerical algorithm completes and reaches it's exit criteria. This typically fails due to the underlying runtime errors from the ODE solver or due to the optimiser algorithm not converging due to reaching the maximum number of iterations allowed; in this paper, we fixed the maximum number of iterations to $10^5$. Here convergence is concerned with how fast the numerical algorithm completes and reaches it's exit criteria, and we quantify this in terms of function evaluations. 

We consider three gradient ascent type methods: vanilla gradient ascent with a Barzilai-Borwein adaptive step size (VGA) \cite{barzilai88}, the Broyden-Fletcher-Goldfarb-Shanno algorithm (commonly known as BFGS), and natural gradient ascent (NGA). If $f$ is the function to be maximised, and $x$ is the parameter of interest, then the VGA iteration is defined by
\begin{align*}
    x_{n+1} = x_n + \alpha_s \nabla f(x_n),
\end{align*}
where $\alpha$ is either a long or short derived from the linear trend of the last two iterates,
\begin{align}
\alpha_{short} &= \frac{ (x_n - x_{n-1}) \cdot (x_n - x_{n-1}) }{  (x_n - x_{n-1}) \cdot (\nabla f(x_n) - \nabla f(x_{n-1}) ) },\\
\alpha_{long} &= \frac{ (x_n - x_{n-1}) \cdot (\nabla f(x_n) - \nabla f(x_{n-1}) ) }{ (\nabla f(x_n) - \nabla f(x_{n-1}) ) \cdot (\nabla f(x_n) - \nabla f(x_{n-1}) ) }.
\end{align}

The next iterate according to BFGS is defined by
\begin{align*}
    x_{n+1} = x_n + \alpha_n \, s_n
\end{align*}
where $s_n$ is obtained by solving
\begin{align*}
    B_n \, s_n = \nabla f(x_n)
\end{align*}
for some $B_n$ which approximates the Hessian matrix of $f$ at $x_n$, and $\alpha_n$ is obtained by solving $\alpha_n = \arg\max_\alpha f(x_n + \alpha \, s_n)$. The next iterate according to NGA is defined by 
\begin{align*}
    x_{n+1} = x_n + \alpha \, s_n
\end{align*}
where $s_n$ obtain by solving
\begin{align*}
    F(x_n) \, s_n = \nabla f(x_n)
\end{align*}
where $F$ is defined as the Fisher Information Matrix associated to $f$ at $x_n$, noting that $f$ must be a likelihood function. In our case, since the ODE is a deterministic function of the parameters, the Fisher Information Matrix only involves the first order sensitivity equations of the system \eqref{eqn:u}--\eqref{eqn:v}. We note that BFGS approximates the Hessian matrix at the current point $x_n$, and uses and adaptive step size, whilst NGA uses the expected value of the Hessian matrix (i.e. the Fisher Information), and we use a constant step size. More precisely, setting $p = (\ta, \delta, u_0, v_0)$ we have
\begin{equation*}
    F_{X_i,Y_i}(\ta, \wg, \wb, \delta, u_0, v_0) = \left(
    \begin{array}{c:c}
        A_u + A_v & B^\top \\
        \hdashline
        B & C
    \end{array}
    \right),
\end{equation*}
where,
\begin{align*}
    [A_u]_i^j &= \wg \left(\partial_{p_i} \exp\left(u\left(\delta t_i;\ta,u_0,v_0\right)\right)\right) \left( \partial_{p_j} \exp\left(u\left(\delta t_i;\ta,u_0,v_0\right) \right)\right), \\
    [A_v]_i^j &= \wb \left(\partial_{p_i} \exp\left(v\left(\delta t_i;\ta,u_0,v_0\right)\right)\right) \left( \partial_{p_j} \exp\left(v\left(\delta t_i;\ta,u_0,v_0\right) \right)\right), \\
    [B]_0^j &= \partial_{p_j} \exp\left(u\left(\delta t_i;\ta,u_0,v_0\right)\right), \\
    [B]_1^j &= \partial_{p_j} \exp\left(v\left(\delta t_i;\ta,u_0,v_0\right)\right),
\end{align*}
and
\begin{align*}
    C = \diag\left( \frac{1}{\wg}  \exp\left(u\left(\delta t_i;\ta,u_0,v_0\right)\right) , \frac{1}{\wb} \exp\left(v\left(\delta t_i;\ta,u_0,v_0\right)\right) \right). 
\end{align*}
Since we are treating $\wb$ and $\wg$ as nuance parameters, we do not need to compute their update via the gradient ascent method, and in turn we only need to compute the inverse of the Fisher Information Matrix corresponding to $\pi$; note that this reduction impacts the three methods, VGA, BFGS and NGA. Although this has little impact for a small problem with only six parameters, it may provide a significant numerical advantage for larger problems, but this is out of the scope of the present study and it is left as future work.

\subsection{Optimisation methods performance}
\begin{figure}[!htb]
    \centering
\includegraphics[width=\linewidth]{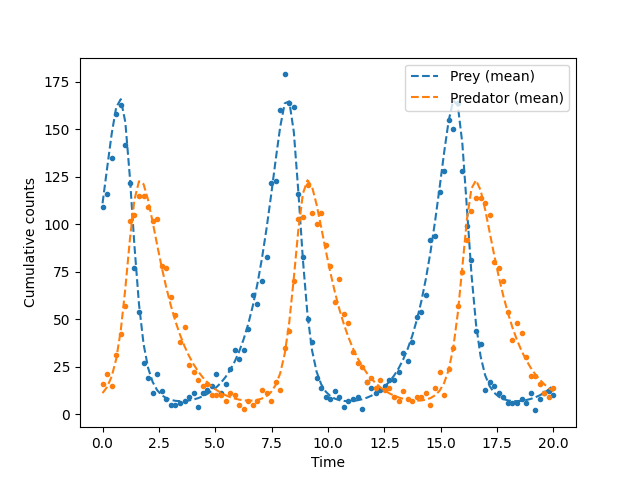}
\caption{Solution to \eqref{eqn:x}--\eqref{eqn:y} with parameters and initial conditions from Table \ref{tab:synth_params}. The simulated observations correspond to samples from a Poisson with mean given by $x(t_i)$ and $y(t_i)$, where $x$ and $y$ solve \eqref{eqn:x}--\eqref{eqn:y}.}\label{fig:synth_data}
\end{figure}

\begin{figure}[!htb]
    \centering
\includegraphics[width=0.9\linewidth]{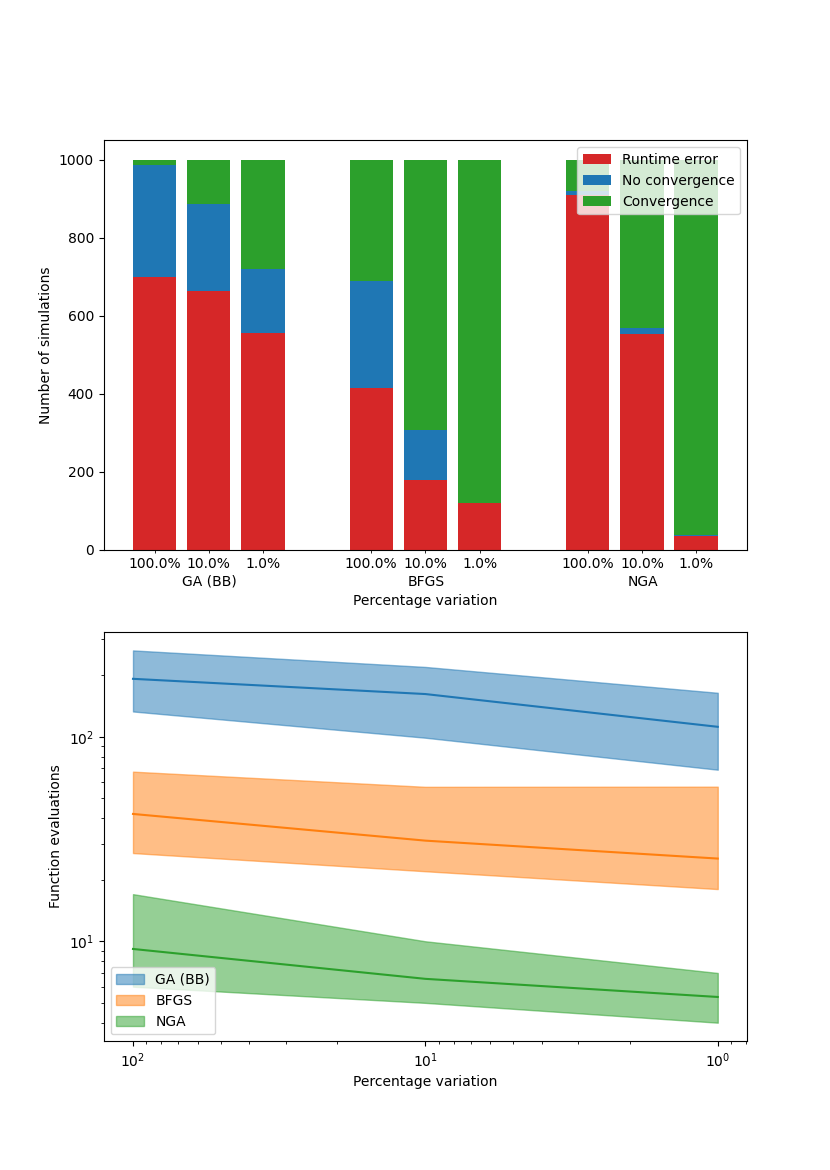}
\caption{Performance of different methods, for different levels of variation of the initial guess around the true value of the parameters and initial conditions, fitting exact observations.}\label{fig:methods_performance}
\end{figure}

\begin{figure}[!htb]
    \centering
\includegraphics[width=0.9\linewidth]{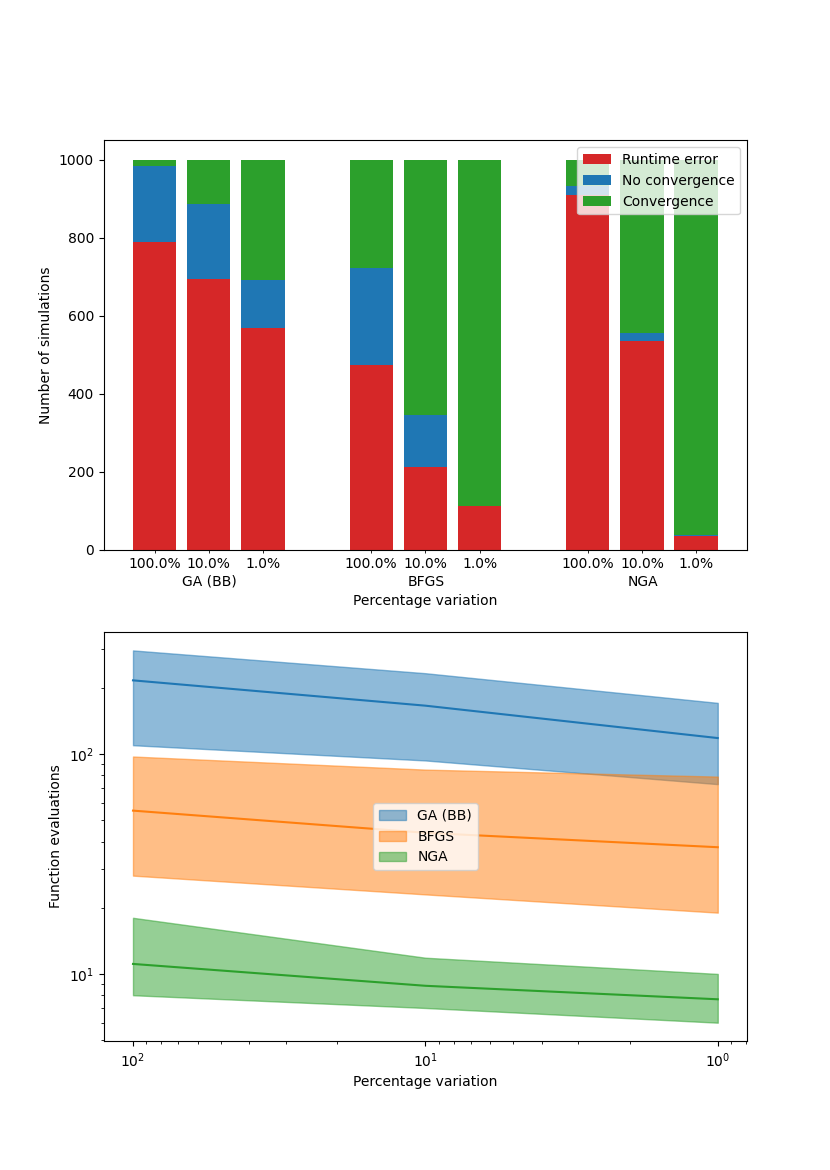}
\caption{Performance of different methods, for different levels of variation of the initial guess around the true value of the parameters and initial conditions, fitting observations with Poisson noise.}\label{fig:poisson_methods_performance}
\end{figure}

To study the performance of the different optimisation methods, we apply them to two datasets: the exact solution of the ODE at different times, and the same solution subject to Poisson noise. We generate the exact solution of the ODE at different times using the parameters and initial conditions depicted in Table \ref{tab:synth_params}, and we display the solutions in Figure \ref{fig:synth_data}. We also display the solution subject to Poisson noise in Figure \ref{fig:synth_data}. We sample initial guesses for the optimisation method from the following distribution
\begin{align*}
    (\ta_0, \log(\delta_0), u_{0,0}, v_{0,0}) \sim N(\mu, \Sigma)
\end{align*}
where
\begin{align*}
    \mu = [\ta, \log(\delta), u_0, v_0]^T, \qquad \text{ and } \qquad \Sigma = \sigma^2 \diag(\ta, \log(\delta), u_0, v_0),
\end{align*}
which come from Table \ref{tab:synth_params}. We note that we sample $\log(\delta)$ rather than $\delta$ since $\delta$ must be strictly positive and we did not want to add extra constraints to the optimisation method. We sample 1000 initial guesses for different values of $\sigma$, namely $\sigma = 1\%$, $10\%$ and $100\%$. Figure \ref{fig:methods_performance} shows that the three methods crash most of the time for inaccurate initial guesses, but BFGS and NGA are more robust than the adaptive gradient ascent in general. More importantly, we can see how NGA takes less iterations to converge than any of the other methods. In Figure \ref{fig:poisson_methods_performance} we confirm that the results persist even in the presence of observational noise in the dataset. 

\begin{table}[!htb]
\caption{Model parameters and initial conditions generating the synthetic data in Figure \ref{fig:synth_data}.}\label{tab:synth_params}
\begin{center}

    \begin{tabular}{|l|c|}%
    \hline \textbf{Parameter} & \textbf{Value} \\ \hline
    \begin{filecontents*}{synthetic_params.csv}
Parameter,Value,latex
a,1.1,$\alpha$
b,0.03,$\beta$
g,0.02,$\gamma$
d,0.9,$\delta$
x0,100,$x_0$
y0,10,$y_0$
tmin,0,$t_{\textrm{min}}$
tmax,20,$t_{\textrm{max}}$
Nobs,100,$N_{\textrm{obs}}$
\end{filecontents*}
\csvreader[head to column names,
                late after line=\\\hline]{synthetic_params.csv}{} {\latex & \Value}
    \end{tabular}

\end{center}
\end{table}

\subsection{NGA performance vs data size}

\begin{figure}[!htb]
    \centering
\includegraphics[width=0.9\linewidth]{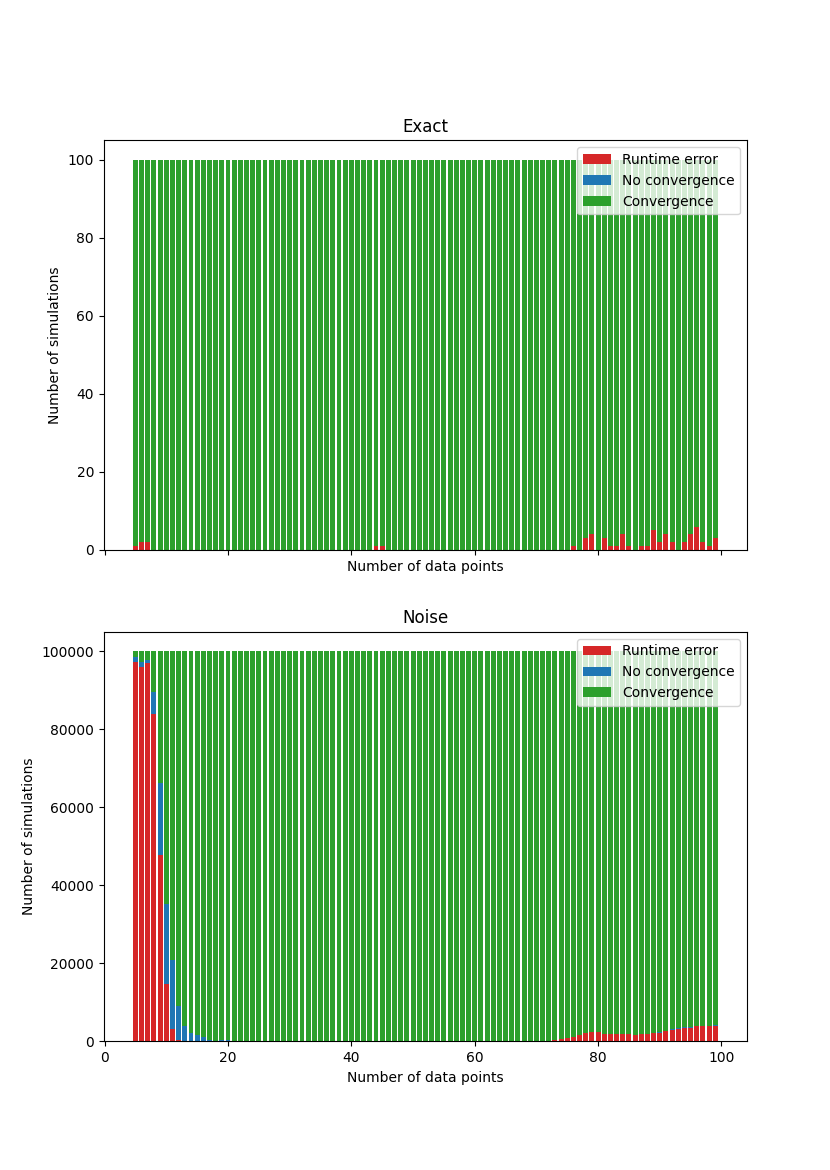}
\caption{Performance of NGA, depending on the number of datapoints used. Top plot depicts exact observations, bottom depicts observations with Poisson noise.}\label{fig:nga_datasize_performance}
\end{figure}

\begin{figure}[!htb]
    \centering
\includegraphics[width=0.8\linewidth]{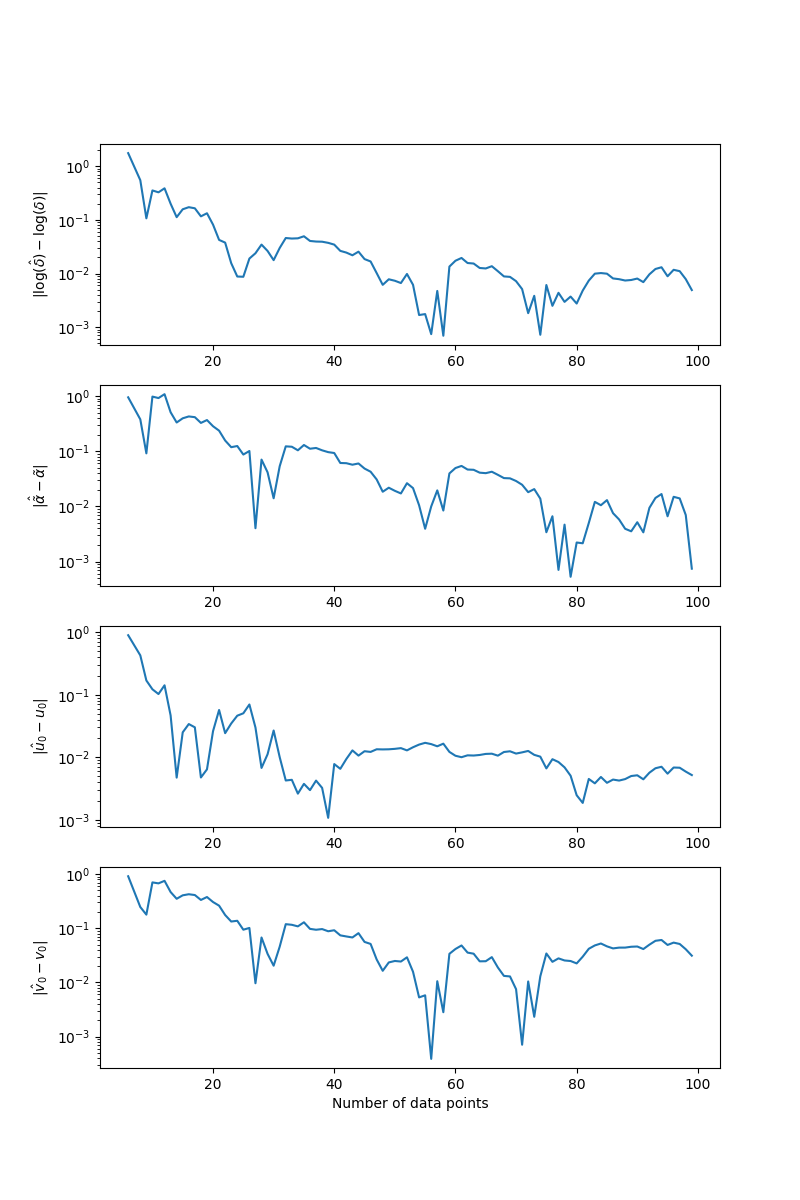}
\caption{Absolute error in the MLE estimates for each parameter as a function of the number of datapoints used for the fitting. Here we consider observations with Poisson noise.}\label{fig:error_datasize}
\end{figure}

We now explore the performance of NGA when we use a small number of observations. We use the same two datasets as before, one with the exact data and one with the Poisson noise. Figure \ref{fig:nga_datasize_performance} shows that the performance is comparable between the exact observations and the observations subject to Poisson noise. The sudden improvement in performance in the region around 10 datapoints corresponds to observing a full cycle of the of the periodic solution; after that point, the periodic solution gives repeated observations of the same state. It is not clear to us why the numerical method starts to have runtime errors when large amounts of data is used. Figure \ref{fig:error_datasize} shows how the absolute error of the MLE decreases with the number of observations.

\section{Discussion}

We explored a computational framework designed to address specific numerical bottlenecks encountered when fitting the classical Lotka-Volterra predator-prey system to sparse data. To mitigate the instability of standard ODE solvers when the optimization trajectory traverses regions of the parameter space that cause the equations to become stiff we exploited intrinsic properties of the model. The method could be generalised to other dynamical systems, as long as we can show that not all the components of the dynamical system become problematic at the same time.  This adaptive approach bypasses the need for very small step sizes. 

Building on this stable numerical foundation, we evaluated the performance of Natural Gradient Ascent (NGA) against standard optimization techniques. There are many arguments to motivate Natural Gradient Ascent: it is a natural metric on the space of distributions, and intuitively in weights the gradient to avoid making progress only in the direction of sensitive components.

The non-dimensionalisation of the dynamical system, that is usually motivated for numerical reasons, also allows us to treat scaling factors as nuisance parameters to reduce dimensionality. The performance of NGA is superior to adaptive gradient ascent with Barzilai-Borwen step, as well as superior to BFGS. We picked these two methods as good baseline methods that are widely used in this setting. Furthermore, NGA consistently converges in less iterations. The performance of NGA is similar both using exact observations and noisy data. We have not explore the performance of an adaptive step for NGA.  

Our main goal was to identify methods that perform well for small datasets. The performance of NGA on sparse data is highly comparable to its performance on larger datasets, and it stabilises quickly as soon as the system becomes observable, and one a full period of the data have been observed. 

\clearpage

\bibliography{ref}

\begin{thebibliography}{6}
\providecommand{\natexlab}[1]{#1}
\providecommand{\url}[1]{\texttt{#1}}
\expandafter\ifx\csname urlstyle\endcsname\relax
  \providecommand{\doi}[1]{doi: #1}\else
  \providecommand{\doi}{doi: \begingroup \urlstyle{rm}\Url}\fi

\bibitem[Barzilai and Borwein(1988)]{barzilai88}
Jonathan Barzilai and Jonathan~M Borwein.
\newblock Two-point step size gradient methods.
\newblock \emph{IMA journal of numerical analysis}, 8\penalty0 (1):\penalty0
  141--148, 1988.

\bibitem[Botev et~al.(2017)Botev, Ritter, and Barber]{martens20}
Aleksandar Botev, Hippolyt Ritter, and David Barber.
\newblock Practical gauss-newton optimisation for deep learning.
\newblock 06 2017.
\newblock \doi{10.48550/arXiv.1706.03662}.
\newblock URL \url{https://arxiv.org/abs/1706.03662v2}.

\bibitem[Campillo-Funollet and Van~Yperen(2026)]{campillo26}
Eduard Campillo-Funollet and James Van~Yperen.
\newblock Comprehensive identifiability analysis and reliable parameter
  estimation for an seir model.
\newblock \emph{arXiv preprint arXiv:2607.09137}, 2026.

\bibitem[Lotka(1926)]{lotka26}
Alfred~J Lotka.
\newblock Elements of physical biology.
\newblock \emph{Science Progress in the Twentieth Century (1919-1933)},
  21\penalty0 (82):\penalty0 341--343, 1926.

\bibitem[Todman et~al.(2023)Todman, Bush, and Hood]{todman23}
Lindsay~C. Todman, Alex Bush, and Amelia~S.C. Hood.
\newblock ‘small data’ for big insights in ecology.
\newblock \emph{Trends in Ecology \&amp; Evolution}, 38, 7 2023.
\newblock \doi{10.1016/j.tree.2023.01.015}.
\newblock URL \url{http://dx.doi.org/10.1016/j.tree.2023.01.015}.

\bibitem[Volterra(1926)]{volterra26}
Vito Volterra.
\newblock Fluctuations in the abundance of a species considered mathematically.
\newblock \emph{Nature}, 118\penalty0 (2972):\penalty0 558--560, 1926.

\end{thebibliography}

\end{document}